\documentclass[11pt,usletter]{article}
\usepackage{jcappub.local}

\pdfoutput=1

\usepackage{graphicx}
\usepackage{hyperref}
\usepackage{color}

\def\ba{\begin{align}}
\def\ea{\end{align}}
\def\be{\begin{equation}}
\def\ee{\end{equation}}
\def\bea{\begin{eqnarray}}
\def\eea{\end{eqnarray}}

\begin{document}
\title{Primordial Magnetic Field from Gravitationally Coupled Electrodynamics in Bouncing Scenario}
\author[a,b]{Jie-Wen Chen}
\author[a,b]{Chong-Huan Li}
\author[a,b]{Yu-Bin Li}
\author[a,b,c]{Mian Zhu}
\affiliation[a]{CAS Key Laboratory for Researches in Galaxies and Cosmology, Department of Astronomy, University of Science and Technology of China, Hefei, Anhui 230026, China}
\affiliation[b]{Department of  Astronomy, Key Laboratory for Researches in Galaxies and Cosmology,  University of Science and Technology of China,   Hefei, Anhui, 230026,  China}
\affiliation[c]{School for Gifted Young,  University of Science and Technology of China,   Hefei, Anhui, 230026,  China}

\emailAdd{chjw@mail.ustc.edu.cn}
\emailAdd{lchh@mail.ustc.edu.cn}
\emailAdd{lyb2166@mail.ustc.edu.cn}
\emailAdd{pb0705@mail.ustc.edu.cn}

\date{\today}

\abstract{
We in this paper study the generation of primordial magnetic field (PMF) in the non-singular bouncing scenario,
through the coupling of the electromagnetic field to gravity.
We adopt an electrodynamic model with a coupling coefficient as a function of the scale factor $a$,
\emph{i.e.} $f=1+(a/a_\star)^{-n}$, with $a_\star$ and $n>0$ being constants.
The result implies that in this mechanism, the power spectrum of PMF today is always blue tilted on large scales from $1$ Mpc to the Hubble length,
and the observational constraints favor the ekpyrotic-bounce scenario.
Furthermore, the back reaction of the energy density of PMF at the bouncing point yields theoretical constraints on the bouncing model.
}

\maketitle

\section{Introduction}
\

Observations have manifested the existence of magnetic field  in the universe,
ranging from the stellar scale ($10^{-5}$ pc) to the cosmological scale ($10^{4}$ Mpc)\cite{Neronov:2010, Taylor:2011}.
In particular, the magnetic field on large scales ($\gtrsim 1$ Mpc) is deemed to be produced in the early universe\cite{Kronberg:1994, Grasso:2001, Dolgov:2001,  Widrow:2002, Kulsrud:2008, Campanelli:2015, Subramanian:2016}, namely, the primordial magnetic field (PMF).
By the recent CMB observations,
the strength of the magnetic field is smaller than a few nano-gauss at the $1$Mpc scale\cite{Planck2015, Zucca:2016}.
Additionally, the $\gamma-$ray detections of the distant blazars imply that the magnetic field should be larger than $10^{-16}$ gauss
on the scales $1-10^{4}$ Mpc \cite{Neronov:2010, Taylor:2011, Tavecchio:2010,  Dolag:2011, Dermer:2011, Batista:2016}.
Note that if alternative methods in data processing are used, the value of the lower limit $10^{-16}$ gauss can be relaxed,
for example, to $10^{-18}$ gauss \cite{Taylor:2011, Dermer:2011, Vovk:2012} and to $10^{-19}$ gauss \cite{Finke:2016}.

The generation of PMF is an unclear and important issue in cosmology.
It is well known that both the standard electrodynamics and the Friedmann universe are conformally invariant,
so that photons cannot be produced in the Friedmann background\cite{Parker:1968}.
Hence, to generate PMF, a mechanism 
which violates the conformal invariance
should be introduced in the early universe \cite{Turner-Widrow, Dolgov:1993, Jerome:2007, Demozzi:2009, Giovannini:2017}.
Typically, the break of conformal invariance can be realized
by introducing a coupling of the electromagnetic field (EMF) to another field
such as gravity\cite{Turner-Widrow, Kunze:2010, Kunze:2012, Jimenez:2010, Jimenez:2011, Qian:2015, Nandi:2017}
and (pseudo-)scalar field\cite{Ratra:1992, Garretson:1992, Jerome:2007, Demozzi:2009, Ferreira:2013, Campanelli:2016, Finelli:2001, Campanelli:2009, Subramanian2010, Membiela2014, Sriramkumar:2015, Chowdhury:2016, Qian:2016, Markkanen:2017, Sharma:2017, Moghaddam:2017}
(and see Ref.\cite{Dolgov:2001, Gasperini:1995, Calzetta:1998, Kandus:2000,  Bamba:2004, Bamba:2007} for more complicated mechanisms).
Most of these mechanisms can be effectively reduced to a model with a time-dependent coupling coefficient of EMF\cite{Demozzi:2009}.
It is worth noting that these models may suffer the strong coupling problem that interaction of the charged particles becomes uncontrollably strong,
and the back reaction problem that the energy of the generated EMF dominates over the background dynamics in the early universe\cite{Demozzi:2009, Ferreira:2013}.
Furthermore, a condition should be satisfied in the model construction, that the standard electrodynamics is recovered in the late stage of universe.

The distribution of PMF today depends on the evolutionary history of the early universe,
so that the observations of PMF can constrain the models of early universe,
if the model of PMF generation is taken.
It is known that inflation is currently the most popular scenario of early universe, and
the bouncing cosmology is also a candidate, alternatively to inflation.
In bouncing cosmology, the universe begins with a contracting phase from a sufficiently large and flat state,
then experiences a nonsingular bouncing phase when the universe is very small,
and finally turns into the expanding phase \cite{Novello:2008, Lehners:2008,  Brandenberger:2011, Brandenberger:2012, Cai:2012, Cai:2014, Battefeld:2015, Brandenberger:2016}.
The PMF generation in bouncing scenario has been studied by Refs. \cite{Membiela2014, Sriramkumar:2015, Chowdhury:2016, Qian:2016, Koley:2016},
and most of them are in the mechanism that the EMF couples to a scalar field.
In this paper, we investigate the PMF generation within a different mechanism that EMF couples to gravity, in bouncing scenario.
A feature of this mechanism is that when the universe is sufficiently flat, for both the initial moments and today,
the standard electrodynamics should be recovered.

The paper is organized as follows.
In Sec. \ref{Magnetogenesis}, we introduce a kind of model of PMF generation from the gravitationally coupled electrodynamics in the bouncing scenario.
In Sec. \ref{bounce model}, we give a brief review of the bouncing cosmology and parameterize the bouncing model.
Afterwards, we analyze the evolution of EMF in the chosen model, in Sec. \ref{Evolution of the Mode}.
We then simplify the power spectrum of PMF today and constrain the bouncing model through the observational data of PMF in Sec. \ref{power spectrum}.
The back reaction of PMF and the constraint of the bouncing model from it are calculated in Sec. \ref{back reaction}.
At last, we conclude with a discussion in Sec. \ref{conclusion}.

In this paper, the natural units with $c=\hbar=1$ is taken.

\section{The Gravitationally Coupled Electrodynamics}\label{Magnetogenesis}
\

We consider the action
\be
\label{action}
S=\int d^4x \sqrt{-g}\left(\mathcal L_{bg}-\frac{1}{4} f^2 F^{\mu \nu} F_{\mu \nu}\right),
\ee
where $\mathcal L_{bg}$ is the Lagrangian of the background,
$F_{\mu \nu}=\partial_\mu A_\nu-\partial_\nu A_\mu$ is the strength tensor of EMF,
and $f$ is the time-dependent coupling coefficient.
In the gravitationally coupled electrodynamics,
$f$ is usually a function of components of the curvature tensors such as $\mathcal R_{\mu \nu}$ and $\mathcal R_{\mu \nu \rho \sigma}$
in the Friedmann universe \cite{Drummond-Hathrell, Turner-Widrow}.
Furthermore, when the universe is sufficiently flat, the standard electrodynamics $f=1$ should be recovered.

The action (\ref{action}) yields the equation of motion of EMF
\be
\label{EoM original}
\partial_{\mu}\left(\sqrt{-g}f^2 F^{\mu\nu}\right)=0.
\ee
To simplify Eq.\eqref{EoM original}, we consider the Coulomb gauge $A^0=\partial_i A^i=0$ in a spatially flat Robertson-Walker background
\be
ds^2=-dt^2+a^2(t)dx^2=a^2(\eta)(-d\eta^2+dx^2).
\ee
Thus the EMF is Fourier expanded by
\begin{align}
A_i(x,\eta)={\mathop \sum \limits_{\sigma=1,2}} \int \frac{d^3 k}{(2\pi)^{3/2}} \epsilon_{i,\sigma}(k)
\big(a_{k, \sigma} A_k(\eta) e^{i k\cdot x} \nonumber \\
+ a^{\dag}_{k, \sigma} A_k^{*}(\eta) e^{-i k\cdot x}\big),
\end{align}
where $\epsilon_{i,\sigma}$ are two orthogonal polarization vectors,
$a_{k,\sigma}$ and $a^{\dag}_{k,\sigma}$ are the annihilation and creation operators satisfying the commutation relation $[a_{k,\sigma}, a^{\dag}_{k',\sigma'}]=\delta_{\sigma, \sigma'} \delta^{(3)}(k-k')$.
The equation of motion \eqref{EoM original} reduces to
\be
\label{EoMAk}
A_k''+2\frac{f'}{f}A_k'+k^2 A_k=0,
\ee
where  $'$  is the derivative with respect to the conformal time $\eta$.
It is convenient to rewrite the equation of motion with respect to the variable $u_k=f A_k$, that
\be
\label{EoMUk}
u_k''+\left( k^2-\frac{f''}{f}\right)u_k=0.
\ee

To investigate the PMF,  a specific formula of $f$ is required.
As is mentioned above, $f$ is a function of the curvature tensors or scalar fields,
namely, a function of the scale factor $a(\eta)$ and its derivatives $a'(\eta)$ and $a''(\eta)$, \emph{etc.}.
For simplicity, the coefficient $f$ can be taken a function of the scale factor only,
and is usually assumed to have a power-law dependance $f \propto a^n(\eta)$ in the bouncing scenario \cite{Membiela2014, Sriramkumar:2015, Chowdhury:2016, Qian:2016}.
In this model, the standard electrodynamics cannot be recovered when the universe is flat $a\rightarrow \infty$.

Here, we extend the power-law model and take
\be
\label{f(a)}
f(a)=1+ \left( \frac{a}{a_{\star}} \right)^{-n},
\ee
where $n$ is a positive constant and $a_\star$ is a characteristic scale factor.
Note that the formula \eqref{f(a)} can serve as an approximation for more complicated models,
\emph{e.g.} the Turner-Widrow model \cite{Drummond-Hathrell, Turner-Widrow}.
In this model, it is clear that when $a\gg a_\star$, for both the initial moments and the late universe,
the standard electrodynamics $f=1$ is recovered.
Furthermore, since $n>0$, the coupling coefficient is always larger than one,
so that the strong coupling problem is absent in this model \cite{Demozzi:2009}.
The value of $a_\star$ can be constrained from observations.
For example, if $a_\star$ were larger than $a_{\text{BBN}}$, the scale factor at the moment of big bang nucleosynthesis (BBN),
the standard electrodynamics at the BBN moment would be violated a lot and the abundances of elements  would be discrepant from the observed results,
hence we can take $a_\star/a_0<a_{\text{BBN}}/a_0\simeq10^{-9}$, where $a_0$ is the scale factor of today.
Additionally, to yield an effective PMF generation, we may expect $a_\star$ to be much larger than the scale factors during the bouncing phase.

From Eq. (\ref{f(a)}), one has
\be
\label{f'/f}
\frac{f'}{f}=\frac{-n \mathcal H}{1+\left( a/a_{\star} \right)^{n}},
\ee
and
\be
\label{f''/f}
\frac{f''}{f}=\frac{n(n+1)\mathcal H^2-n\frac{a''}{a}}{1+\left( a/a_{\star} \right)^{n}}=\frac{\left(n^2+\frac{1+3w}{2}n\right)\mathcal H^2}{1+\left( a/a_{\star} \right)^{n}},
\ee
where $\mathcal H\equiv a'/a$ is the comoving Hubble parameter and $w\equiv p/\rho$ is the equation-of-state(EoS) parameter.
Eqs. \eqref{f'/f} and \eqref{f''/f} imply that $|f'/f| \ll |\mathcal H|$ and $|f''/f| \ll |\mathcal H^2|$ for $a\gg a_\star$,
and $|f'/f| \sim |\mathcal H|$ and $|f''/f| \sim |\mathcal H^2|$ for $a \leq a_\star$

\begin{figure} [h]
\begin{center}
\includegraphics[width=0.8\textwidth]{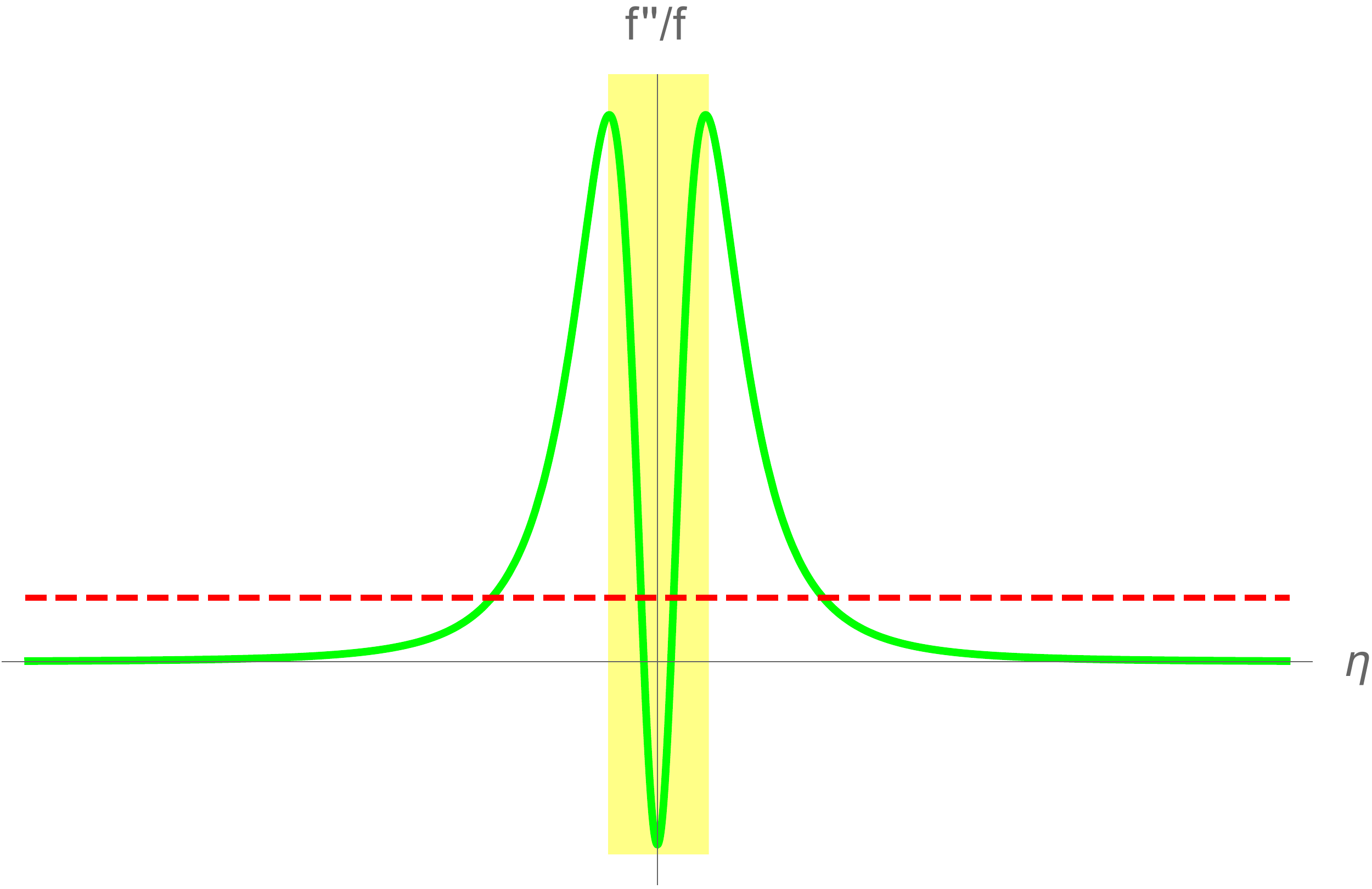}
\caption{\label{Fig f''/f}
A sketch of the evolution of $f''/f$ in the non-singular bouncing cosmology (green solid).
The red dashed line is for the scale with comoving wavenumber $k$.
The yellow area corresponds to the bouncing phase.
}
\end{center}
\end{figure}

The evolution  of $f''/f$ is sketched in Fig. \ref{Fig f''/f}.
In the contracting stage, $f''/f$ increases from 0 according to \eqref{f''/f}, since the $|\mathcal H|$ increases from 0.
During the bouncing phase, the value of $f''/f$ should decrease soon and be negative at the bouncing point, due to $\mathcal H=0$ and $a''/a>0$,
and then increase after the bouncing point symmetrically.
In the expanding universe, $f''/f$ should decrease to 0.
From Fig. \ref{Fig f''/f}, it is seen that a long-wavelength mode
 generally crosses the characteristic scale for 4 times,
wherein the first crossing occurs in the contracting stage;
the second and third ones happen around the bounce point;
and the forth occurs during the expanding stage.
A detailed analysis on the evolution of the mode $A_k$ will be given the Sec. \ref{Evolution of the Mode}.

\section{Bouncing Background}\label{bounce model}
\

The bouncing scenario can be achieved by various mechanisms, e.g.
modifying the gravitational action such as Horava-Lifshitz gravity bounce\cite{Brandenberger:2009yt, Cai:2009in, Gao:2009wn}
and $f(T)$ teleparallel bounce\cite{Cai:2011tc, deHaro:2012zt, Cai:2015emx},
and introducing a Null Energy Condition (NEC) violating field such as quintom bounce\cite{Cai:2007zv, Cai:2008ed},
ghost condensate bounce \cite{Lin:2010pf},
Lee-Wick bounce\cite{Cai:2008qw}, Galileon bounce\cite{Qiu:2011cy, Easson:2011zy} and fermionic bounce\cite{Alexander:2014eva, Alexander:2014uaa}, \emph{etc.}.
In the present paper, we do not concern a specific mechanism of bouncing, and take a general discussion in the frame of general relativity.

The bouncing cosmology is divided to three stages, and each one is modeled as follows. \\
i) During $t<t_{-}$ the universe is in the contracting stage, where the subscript ``-" denotes the end of the contracting phase.
The EoS parameter of this stage $w$ is assumed to be a constant for simplicity.
Thus the scale factor and Hubble parameter during the contracting stage follow
\be
a^{3(1+w)}H^2=a_{-}^{3(1+w)}H_{-}^2,
\ee
and the contracting stage can be fixed by three parameters $a_{-}$, $H_{-}$ and $w$.
In specific, $w=0$ corresponds to the matter-bounce model $w=0$\cite{Brandenberger:2009yt, Brandenberger:2012, Cai:2011tc, Lin:2010pf, Cai:2013kja} and $w\gg 1$ to the ekpyrotic-bounce model\cite{Lehners:2008}.
Note that the bouncing models with $w<1$ generally suffer the Belinsky-Khalatnikov-Lifshitz instability,
namely, the fluctuation of anisotropy dominates over the background dynamics\cite{BKL}, but we do not consider this problem here.
Furthermore, we expect that the universe should be initially flat, $\mathcal R_{\alpha \beta \gamma \delta} \rightarrow 0$ for $a\rightarrow \infty$,
which requires $w>-1/3$.\\
ii) During $t_{-}<t<t_{+}$ the universe is in the bouncing stage, where the subscript ``+" denotes the end of the bouncing phase,
and $t=0$ is the bouncing point.
During this stage, the Hubble parameter is described by
\be
\label{H(t)bouncing}
H=\Upsilon t,
\ee
where $\Upsilon$ is a constant, so that $t_{-}=H_{-}/\Upsilon$ and $t_{+}=H_{+}/\Upsilon$.
From Eq.\eqref{H(t)bouncing}, the scale factor is given by
\be
\label{a(t)bouncing}
a(\eta)=a_b e^{\frac{\Upsilon t^2}{2}},
\ee
where $a_b$ is the scale factor at the bouncing point.
The bouncing phase described by \eqref{H(t)bouncing} and \eqref{a(t)bouncing}
can be exactly achieved by a scalar field with a Horndeski-type, non-standard kinetic term and
a negative exponential potential\cite{Cai:2012, Cai:2014},
and it can also serve as a general approximation of most bouncing phases.\\
iii) When $t>t_{+}$ the universe is in the expanding stage, described by the classical big bang model.
For simplicity, we only consider the radiation dominated(RD) stage,
since during the whole matter dominated and dark energy dominated stages,
the scale factor only changes three orders of magnitude.
Hence, one has
\be
\label{RD}
a^2 H=a_{+}^2 H_{+}=a_0^2 H_0
\ee
during the expanding stage, where $a_0$ and $H_0$ are the scale factor and Hubble parameter  today.

Given above, the bouncing model is described by four parameters: $w$, $H_{-}$, $\Upsilon$ and $H_{+}$.
The evolution of the EMF in the bouncing model will be given in the next section.

\section{Evolution of EMF} \label{Evolution of the Mode}
\

During the initial and late moments,
the short-wavelength condition $k^2 \gg f''/f$ is satisfied,
so the solution of Eqs. (\ref{EoMAk}) is
\be
\label{Ak-subhorizon}
A_k = \frac{c_1 e^{-i k \eta}}{f}+\frac{c_2 e^{i k \eta}}{f},
\ee
where $c_1$ and $c_2$ are time-independent coefficients.
It is worth noting that around the bouncing point, $k^2>f''/f$ is also satisfied,
but EMF does not follow Eq.(\ref{Ak-subhorizon}),
due to the non-vanishing effective potential $f''/f<0$.
For the wavelengthes much larger than the time scale of the bouncing phase,
i.e. $k\ll \frac{1}{\eta_{+}-\eta_{-}} \sim \frac{a_b \Upsilon}{H_{+}-H_{-}}$,
the term $k^2 A_k$ in  \eqref{EoMAk} can be negligible when integrating the equation over the bouncing phase,
so that these modes should behave as $k^2\ll f''/f$ during the bouncing phase.
This result is similar to the unchanged behavior of the curvature perturbation around the bouncing point\cite{Battarra:2014, Quintin:2015}.
For the long-wavelength modes $k^2\ll f''/f$, the solution of Eq.(\ref{EoMAk}) is
\be
\label{Ak-superhorizon}
A_k(\eta)=A_k(\eta_i)+A_k'(\eta_i) f^2(\eta_i) \int^{\eta}_{\eta_i} \frac{d\tilde{\eta}}{f^2(\tilde{\eta})},
\ee
where $\eta_i$ is an arbitrary moment of $k^2\ll f''/f$, and the derivative of the mode follows
\be
\label{Ak'-superhorizon}
 A_k'(\eta) f^2(\eta)=A_k'(\eta_i) f^2(\eta_i)= {\emph{const}}.
\ee
During a stage with a constant EoS parameter $w$, for both the contracting and expanding phases, the integral in Eq. \eqref{Ak-superhorizon} reduces to
\be
\label{A_k w}
\int \frac{d\eta}{f^2}= \int \frac{da}{a^2 H f^2} \propto
 \begin{cases}
a^{\frac{3w+1}{2}}, &  a\gg a_\star\\
a^{\frac{3w+1}{2}+2n}, & a\ll a_\star
 \end{cases}~.
\ee
Thus the constant term $A_k(\eta_i)$ dominates the right hand side of Eq. \eqref{Ak-superhorizon} during the contracting stage with $w>-1/3$,
and the integral term can be considered  in the expanding stage.
During the bouncing phase, as the time interval $\eta_{+}-\eta_{-}$ is very short for the concerned modes,
and the integrand $1/f^2$ is small, the integral term in \eqref{Ak-superhorizon} is negligible.
Therefore, the EMF is amplified mainly during the RD era after the bouncing phase.

\begin{figure*}
 \centering
\includegraphics[scale=0.5]{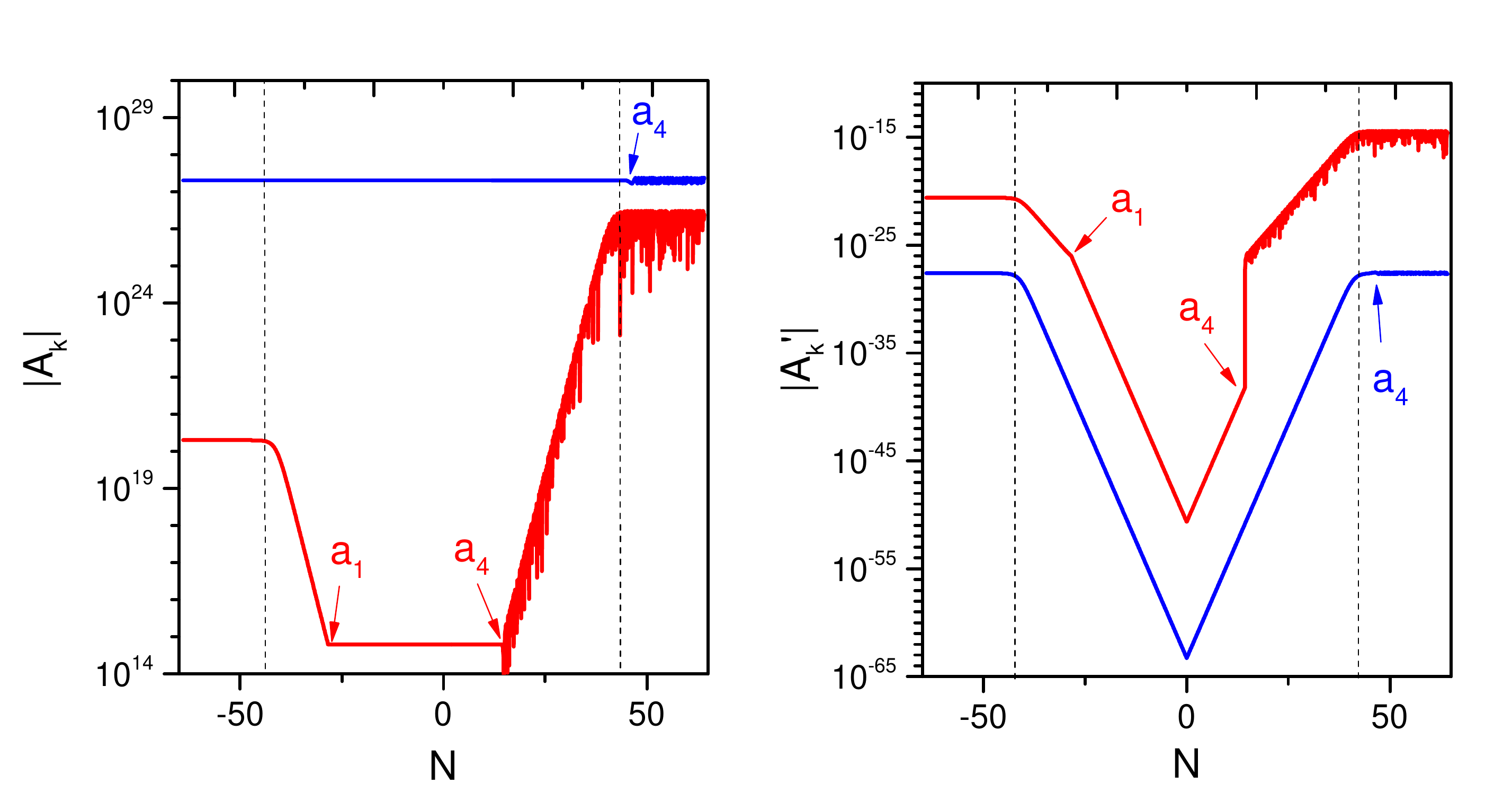}
\caption{The left panel sketches the evolutions of $|A_k|$ and the right panel sketches that of $|A_k'|$.
The blue solid curve is for a long-wavelength mode with $k= 10^6 \mathcal H_0$, the red solid one for a short-wavelength one with $k=10^{20}\mathcal H_0$,
and the dashed line labels $a_\star$.
The horizontal axis is the e-folding number $N=\ln(a/a_b)$, and the sign of $N$ is taken to be minus before the bouncing point.
In this case, we choose the model parameters as $w=0$, $|H_{-}|=H_{+}=10^{-7} M_p$, $\Upsilon=5\times 10^{-7} M_p^2$, $n=1$ and $a_\star=10^{-10} a_0$.
\label{Ak evolution}
}
\end{figure*}

Now we give a brief summary of the evolution of $A_k$ over all stages in the bouncing scenario.
Initially $f=1$, the Bunch-Davies vacuum is taken, so that $c_1=1$, $c_2=0$, and the mode is $A_k=e^{-i k \eta}/\sqrt{2k}$.
From  Eq.(\ref{Ak-subhorizon}), one obtains the mode at the first time of $k^2=f''/f$ in the contracting stage
\be
A_k(\eta_1)=\frac{e^{-i k \eta_1}}{f_1 \sqrt{2k}},\ \ \ \ \ \ \ \ A_k'(\eta_1)=\left(-\frac{f_1'}{f_1}-ik\right) A_k(\eta_1),
\ee
where the subscript ``$l$" $(l=1, 2, 3, 4)$ denotes the $l^{th}$ time  when $k^2=f''/f$.
The value of $a_1$, for a fixed wavenumber $k$, is obtained by
\be
\label{crossing 1}
k^2=\frac{\left(n^2+\frac{1+3w}{2}n\right)\mathcal H_1^2}{1+(a_1/a_\star)^{n}},
\ee
where $a_1>a_{-}$ and $\mathcal H_1^2=\mathcal H_{-}^2 a_{-}^{1+3w}/a_{1}^{1+3w}$.
As discussed  above, $A_k$ is frozen from $\eta_1$ until the end of the bouncing phase, hence
\be
\label{A_k2}
A_k(\eta_{+})\simeq A_k(\eta_1),\ \ \ \ \ \ \ \ A_k'(\eta_{+})=A_k'(\eta_1) \frac{f_1^2}{f_{+}^2}.
\ee
During the RD era, $A_k$ is amplified, and the mode at $\eta_4$ is
\begin{align}
\label{A_k4}
A_k(\eta_4) = A_k(\eta_{1})+A_k'(\eta_{1}) f_{1}^2 \int^{\eta_4}_{\eta_{+}}\frac{d\eta}{f^2}, \ \ \ \nonumber \\
A_k'(\eta_4)=A_k'(\eta_{1}) \frac{f_{1}^2}{f_4^2}. ~ ~ ~ ~ ~ ~ ~ ~ ~ ~ ~ ~
\end{align}
The value of $a_4$ is given by
\be
\label{crossing 4}
k^2=\frac{\left(n^2+n\right)\mathcal H_4^2}{1+(a_4/a_\star)^{n}},
\ee
with $a_4>a_{+}$  and $\mathcal H_4^2=\mathcal H_{+}^2 a_{+}^{2}/a_{4}^{2}$.
After  $\eta_4$, the mode evolves as (\ref{Ak-subhorizon}),
or rewritten as
\be
A_k=\frac{e_1\cos k(\eta-\eta_4)+e_2\sin k(\eta-\eta_4)}{f},
\ee
where the coefficients $e_1$ and $e_2$ are fixed by the conjunction condition at $\eta_4$
\be
\label{e1e2}
e_1=f_4 A_k(\eta_4), \ \ \ \ \ \ \ \ \ \ \ e_2=\frac{f_4}{k}(A_k'(\eta_4)+A_k(\eta_4)\frac{f_4'}{f_4}).
\ee
Despite the oscillation, the profile of the mode after $\eta_4$ is $|A_k|\simeq \frac{\sqrt{|e_1|^2+|e_2|^2}}{f}$.
Hence, the mode today is
\begin{align}
\label{Ak0}
A_k(\eta_0) = e_1\cos k(\eta_0-\eta_4)+e_2\sin k(\eta_0-\eta_4) \nonumber \\
\simeq  e_1\cos \frac{k}{\mathcal H_0}+e_2\sin \frac{k}{\mathcal H_0},~ ~ ~ ~ ~ ~ ~ ~ ~ ~ ~ ~
\end{align}
and the profile is $|A_k(\eta_0)|\simeq \sqrt{|e_1|^2+|e_2|^2}$.
The evolutions of $A_k$ and its time derivative $A_k'$ are sketched in Fig. \ref{Ak evolution},
for a long-wavelength mode and a short-wavelength one respectively.

\section{Power spectrum}\label{power spectrum}
The power spectrum of PMF today can be calculated by
\be
\label{Power}
P_{B} (k, \eta_0)=(B_\lambda)^2=\frac{k^5 |A_k(\eta_0)|^2}{2\pi^2 a_0^4},
\ee
where $B_\lambda$ is the mean strength of the magnetic field on the physical length scale $\lambda\simeq a_0/k$.
We focus the interest on the magnetic field on scales from $1$Mpc to the Hubble length,
which corresponds to comoving wavenumber from $\mathcal H_0$ to $\mathcal O(10^4)~\mathcal H_0$.
According to Eq. \eqref{crossing 4}, $a_4\leq a_\star$ implies $k \gtrsim \frac{a_0}{a_\star} \mathcal H_0> 10^9 \mathcal H_0$,
so that all the modes on the concerned scales satisfy $a_4 \gg a_\star$ and $f_4\simeq 1$, hence
\be
\label{f4=1}
 k\simeq \frac{\mathcal H_4}{(a_4/a_\star)^{n/2}} \ll \mathcal H_4, \ \ \ \
\int^{\eta_4}_{\eta_{+}}\frac{d\eta}{f^2} \simeq \eta_4 \simeq \mathcal H_4^{-1}.
\ee
Eq. \eqref{crossing 4} reduces to
\be
a_4=a_\star \left( \frac{(n^2+n)a_{+}^4 H_{+}^2}{a_\star^2} \right)^{\frac{1}{n+2}} k^{-\frac{2}{n+2}},
\ee
and the coefficients in \eqref{e1e2} are
\be
\label{e3e4}
e_1\simeq A_k(\eta_4), \ \ \ \ \ \ \ e_2 \simeq \frac{A_k'(\eta_4)}{k} =\frac{A_k'(\eta_1)}{k} f_1^2.
\ee

We still need to specify the value of $f_1$.
For the case $f_1 \simeq 1$, i.e. $a_1 \gg a_\star$,
$k\simeq \frac{\mathcal H_1}{(a_1/a_\star)^{n/2}}$ is much larger than $\frac{f_1'}{f_1} \simeq \frac{\mathcal H_1}{(a_1/a_\star)^{n}}$
according to Eqs. \eqref{f'/f} and  \eqref{crossing 1},
thus $|A_k'(\eta_1)|\simeq k |A_k(\eta_1)|$ and
\begin{align}
e_1\simeq A_k(\eta_4)\simeq A_k(\eta_1)-ik A_k(\eta_1) \mathcal H_4^{-1} \simeq A_k(\eta_1), \nonumber \\
e_2 \simeq A_k(\eta_1), \ \ \  \ \ \ \ \ \ \  \ \ \ \ \ \ \ \ \ \ \ \ \ \ \ \ \ \ \  \   \nonumber
\end{align}
by Eqs. \eqref{A_k4}, \eqref{f4=1} and \eqref{e3e4}.
Thus today's mode is $|A_k(\eta_0)|\simeq \sqrt{|e_1|^2+|e_2|^2} \sim  |A_k(\eta_1)|$,
and the strength  by \eqref{Power} is
\[
B_\lambda\simeq \frac{k^2}{2\pi a_0^2}\sim  \left(10^{-57} \text{G}\right) \left(\frac{k}{10^4 \mathcal H_0} \right)^2,
\]
which is much smaller than the lower limit $10^{-19} \text{G}$ on the concerned scales \cite{Finke:2016}.
Therefore, the case $f_1\simeq 1$ will not be considered in the following.

\begin{figure}
\begin{center}
\includegraphics[scale=0.5]{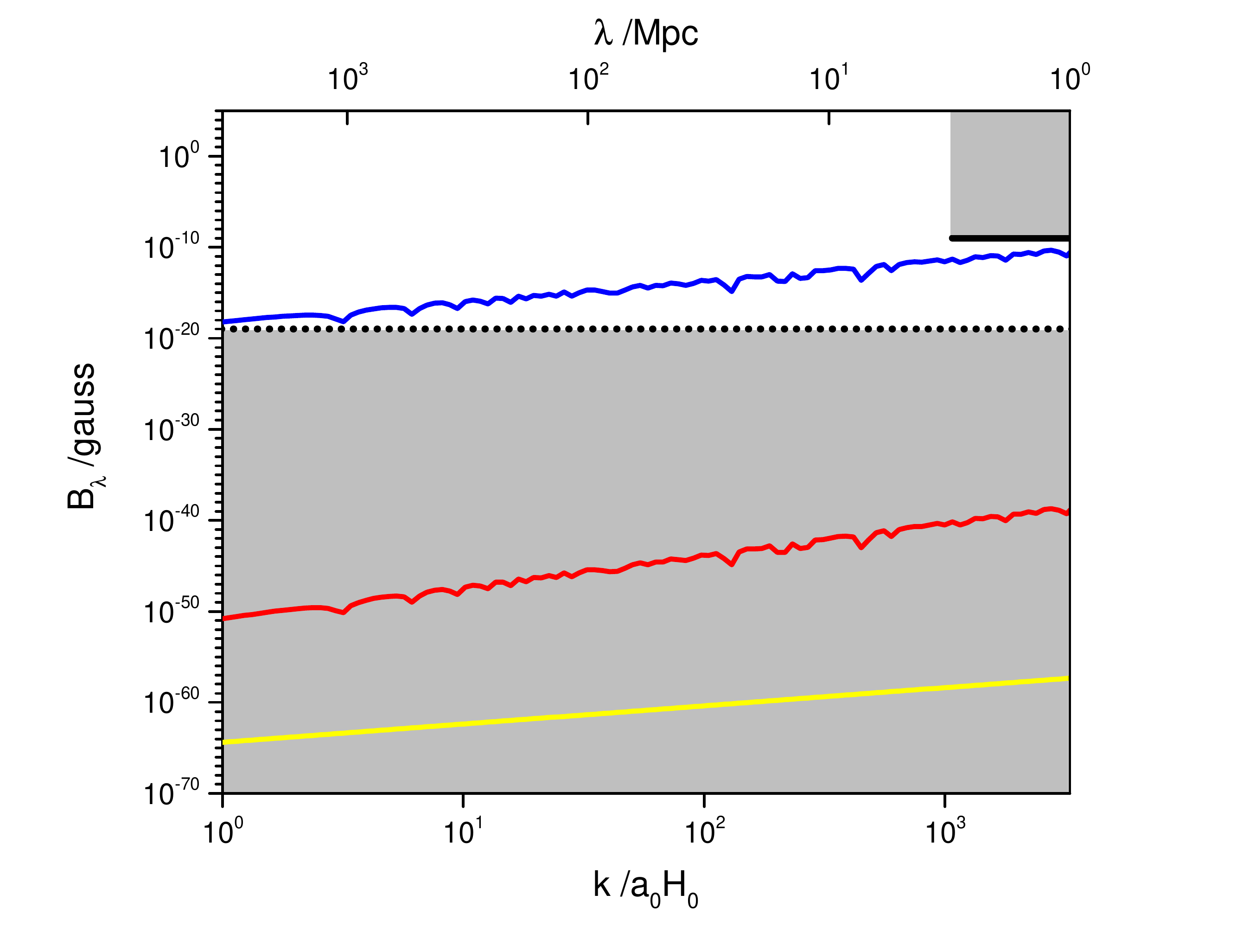}
\caption{The large-scale $B_\lambda$ for various $w$, at $|H_{-}|=H_{+}=10^{-7}M_p$, $\Upsilon=5\times 10^{-7}$, $n=3$ and $a_\star=10^{-10}$.
The yellow curve is for $w=0$, the red for $w=1$ and the blue for $w=7$.
The black solid line is the upper bound constrained by CMB observations\cite{Planck2015, Zucca:2016},
the dotted line is the lower bound from \cite{ Finke:2016}, and the gray regimes are excluded by observations.
\label{B-w}
}
\end{center}
\end{figure}

For $f_1 \gg 1$, \emph{i.e.} $a_1 \ll a_\star$,
one has  $k\sim \frac{f_1'}{f_1} \sim \mathcal H_1$ by \eqref{f'/f}, so
\be
\label{e5e6}
e_1\simeq A_k(\eta_4)\simeq A_k(\eta_1)-ik A_k(\eta_1) f_1^2 \mathcal H_4^{-1},
\ee
by \eqref{A_k4}.
To  amplify $A_k$ significantly from $\eta_1$ to $\eta_4$, $k f_1^2 \mathcal H_4^{-1} \gg 1$ is required,
so that $A_k'(\eta_4) \simeq \mathcal H_4 A_k(\eta_4)$ according to \eqref{f4=1},
and $e_2 \simeq \frac{\mathcal H_4}{k} A_k(\eta_4) \gg e_1$.
Hence
\be
|A_k(\eta_0)| \simeq |e_2|\simeq \frac{\mathcal H_4}{k} k |A_k(\eta_1)| f_1^2 \mathcal H_4^{-1}= \frac{f_1}{\sqrt{2k} },
\ee
where
\be
f_1 \simeq \left(\frac{a_1}{a_\star} \right)^{-n} \simeq \left( \frac{(n^2+\frac{1+3w}{2}n) a_{-}^{1+3w} a_{-}^2 H_{-}^2}{a_\star^{1+3w}} \right)^{-\frac{n}{1+3w}} k^{\frac{2n}{1+3w}},
\ee
by \eqref{crossing 1}.
From Eq. \eqref{Power}, the resulting strength is
\begin{align}
\label{B}
B_\lambda \simeq 10^{-65} \text{G} \left(n^2+\frac{1+3w}{2}n \right)^{-\frac{n}{1+3w}}\left(\frac{k}{\mathcal H_0} \right)^{2+\frac{2n}{1+3w}} \nonumber \\
\left( \frac{a_\star}{a_0} \right)^n \left( \frac{H_{+}}{|H_{-}|}\right)^{\frac{2n}{1+3w}} \left( \frac{H_+}{H_{0}}\right)^{\frac{n}{2}-\frac{n}{1+3w}} \nonumber \\
e^{\left(\frac{n}{2}+\frac{2n}{1+3w} \right)\frac{H_{+}^2-H_{-}^2}{2\Upsilon}}.
\end{align}
It is clear that  $B_\lambda \propto k^{2+\frac{2n}{1+3w}}$ is always blue tilted, with an index $n_B=2+\frac{2n}{1+3w}>2$.
Combining the upper limit $B_{\lambda}<10^{-9}$ G from CMB observations and the lower limit $B_{\lambda}>10^{-19}$G from the $\gamma$-ray observations,
the index is constrained by $n_B<2.25$ (see Fig. \ref{B-w}).
From this aspect, the observations favor the bouncing model with a large $w$.

In addition,
to generate PMF with strength larger than $10^{-19}$ G on the concerned scales,
one needs $\left( \frac{H_{+}}{|H_{-}|}\right)^{\frac{2n}{1+3w}} \left( \frac{H_+}{H_{0}}\right)^{\frac{n}{2}-\frac{n}{1+3w}}
e^{\left(\frac{n}{2}+\frac{2n}{1+3w} \right)\frac{H_{+}^2-H_{-}^2}{2\Upsilon}} \gg 1$ according to Eq. \eqref{B}.
Hence, at least one of the following three conditions should be satisfied:
(i) $\left(\frac{H_{+}}{|H_{-}|}\right)\gg1$, (ii) $w>1/3$   and  (iii) $\frac{H_{+}^2-H_{-}^2}{\Upsilon} \gg 1$.

Given above, from consideration of both the amplitude and the index $n_B$,
the observations favor the ekpyrotic-bounce scenario,
and this is illustrated in Fig. \ref{B-w}.

\section{Back reaction}\label{back reaction}
\

The energy densities of  electric and  magnetic fields are respectively\cite{Jerome:2007, Membiela2014}
\be
\label{density}
\rho_E=\int \frac{dk}{k} \frac{f^2 k^3 |A_k'|^2}{2\pi^2 a^4} \ \ \ \ \text{and} \ \ \ \ \rho_B=\int \frac{dk}{k} \frac{f^2 k^5 |A_k|^2}{2\pi^2 a^4},
\ee
 and the total density of EMF is $\rho_{EM}=\rho_E+\rho_B$.
During the bouncing phase, the mode and its derivative evolve as $A_k\simeq A_k(\eta_1)$ and $A_k'\simeq k A_k(\eta_1) f_1^2/f^2$,
by the discussions in Secs. \ref{Evolution of the Mode} and \ref{power spectrum}.
Thus Eq. \eqref{density} yields
\be
\label{EM density}
\rho_{EM}=(f^2+f_1^2)\int \frac{dk}{k} \frac{k^5 |A_k(\eta_1)|^2}{2\pi^2 a^4}\simeq \frac{f^2}{4\pi^2 a^4}\int^{k_c}_{0} dk \frac{k^3}{f_1^2},
\ee
where the upper limit is taken by $k_c=|\mathcal H_{-}|$, i.e.  only the classical modes which have exited the scale $\sqrt{f''/f}$  are considered. 
It is seen that $\rho_{EM} \propto f^2/a^4 \propto a^{-2n-4}$ during the bouncing phase,
behaving as a fluid with EoS parameter $w_{EM}=\frac{2n+1}{3}$.
Eq. \eqref{EM density} can  approximately reduce to
\begin{align}
\label{EM density 2}
\rho_{EM}\simeq \frac{f^2 k_\star^4}{4\pi^2 a^4}
\left( \frac{1}{4}+
\begin{cases}
\ln\left( \frac{k_c}{k_\star} \right), & n=1+3w\\
\frac{1}{\frac{4n}{1+3w}-4}, & n>1+3w\\
\frac{1}{4-\frac{4n}{1+3w}} \left( \frac{k_c}{k_\star} \right)^{4-\frac{4n}{1+3w}}, & n<1+3w
\end{cases} \right),
\end{align}
where $k_\star = |\mathcal H_{-}| (a_{-}/a_\star)^{(1+3w)/2}$,
because
for $k<k_\star$, one has $a_1>a_\star$ and $f_1=1$, and the integrand in \eqref{EM density} is $\frac{k^3}{f_1^2} = k^3$;
and for $k>k_\star$, the integrand is  $\frac{k^3}{f_1^2} \propto k^{3-\frac{4n}{1+3w}}$.


The energy density of the generated EM field  should not dominate the evolution of the universe,
which can give  theoretical constraints to the bouncing model.
For simplicity, we only consider the back reaction during the bouncing phase.
At the beginning of the bouncing phase, the density of the EM field should be smaller than the background density $\rho_{EM}(\eta_{-})<3M_p^2 H_{-}^2$.
For the same reason, at the end of the bouncing phase, one has $\rho_{EM}(\eta_{+})<3M_p^2 H_{+}^2$.
Furthermore, around the non-singular bouncing point, the background density $\bar \rho$ vanishes,
and the pressure of the background is negative $\bar p(\eta=0)=-2M_p^2 \frac{\ddot a}{a^2}=-2\Upsilon M_p^2$ according to the Friedmann equation,
so the null energy condition is violated $\bar \rho+\bar p <0$.
Considering the back reaction of EMF, including the density $\rho_{EM}$ and the pressure $p_{EM}=w_{EM} \rho_{EM}$ , if the null energy condition is recovered, the nonsingular-bouncing phase wound no longer hold,
so one has another constraint $\rho_{EM} (\eta=0)<\frac{3}{2+n}\Upsilon M_p^2$.

For example, we choose the model with $w=7$, $n=3$, $|H_{-}|=H_{+}=10^{-7}M_p$ and $a_\star=10^{-10}$,
and take $\Upsilon$ as a parameter.
Then from \eqref{EM density 2}, one can calculate that $\rho_{EM}(\eta_{-})\simeq \frac{H_{-}^4}{16\pi^2}$,
automatically satisfying the condition $\rho_{EM}(\eta_{-})<3M_p^2 H_{-}^2$.
In addition, as the EoS parameter of EMF is $w_{EM}=\frac{1+2n}{3}$ during the bouncing phase,
the energy density at $\eta_{+}$ is $\rho_{EM}(\eta_{+})=e^{(n+2)\frac{H_{-}^2-H_{+}^2}{\Upsilon}}\rho_{EM}(\eta_{-})=\rho_{EM}(\eta_{-})$,
also fulfilling the condition $\rho_{EM}(\eta_{+})<3M_p^2 H_{+}^2$.
Furthermore, at the bouncing point, the condition
$\rho_{EM} (\eta=0)=e^{(n+2)\frac{H_{-}^2}{\Upsilon}}\rho_{EM}(\eta_{-})<\frac{3}{2+n}\Upsilon M_p^2$
yields a constraint $\Upsilon > 1.4 \times 10^{-15} M_p^2$.
Thus the bouncing model is constrained.

\section{Conclusion}\label{conclusion}
\

We have investigated the PMF generation in the non-singular bouncing scenario,
through the coupling of the electromagnetic field to gravity.
In this mechanism, when the universe is sufficiently flat, the standard electrodynamics should be recovered,
for both the initial moments and the present day in the bouncing universe.
In specific, the time-dependent coupling coefficient is assumed to be $f=1+\left(\frac{a}{a_\star}\right)^{-n}$
with $n>0$, so that the strong coupling problem is absent in this model.
Additionally, the bouncing cosmology is not studied based on a specific model,
but by a generic parametrization in the frame of general relativity.

The evolution of EMF in the bouncing universe is analysed with some approximations.
We concentrate the interest on the large-scale PMF ranging from $1$Mpc to the Hubble length.
We find that the power spectrum of PMF today on these scales is always blue tilted in this mechanism.
To obtain PMF in accordance to the observational constraints,
especially a nearly scale-invariant power spectrum $n_B<2.25$,
the universe should have a large EoS parameter in the contracting stage,
namely, the ekpyrotic-bounce scenario is favored.
Moreover, the energy density and pressure of the generated EMF around the bouncing point is calculated.
The requirement that EMF should not dominate the background evolution
yields additional constraints on the bouncing model.
\\
\\

\section*{Acknowledgments}
We are grateful to Prof. Y.~F.~Cai for valuable comments.
Y.~F.~Cai and M.~Zhu are supported in part by the Chinese National Youth
Thousand Talents Program and by the USTC start-up funding (Grant No. KY2030000049) and by the National Natural Science Foundation of China (Grant No. 11421303).
Part of numerical computations are operated on the computer cluster LINDA in the particle cosmology group at USTC.

\end{document}